# Data monitoring committees for clinical trials evaluating treatments of COVID-19

Tobias Mütze[1], Tim Friede[2,3]

[1]Statistical Methodology, Novartis Pharma AG, Basel, Switzerland
[2]Department of Medical Statistics, University Medical Center Göttingen, Göttingen, Germany
[3]DZHK (German Center for Cardiovascular Research), partner site Göttingen, Göttingen, Germany

**Abstract.** The first cases of coronavirus disease 2019 (COVID-19) were reported in December 2019 and the outbreak of SARS-CoV-2 was declared a pandemic in March 2020 by the World Health Organization. This sparked a plethora of investigations into diagnostics and vaccination for SARS-CoV-2, as well as treatments for COVID-19. Since COVID-19 is a severe disease associated with a high mortality, clinical trials in this disease should be monitored by a data monitoring committee (DMC), also known as data safety monitoring board (DSMB). DMCs in this indication face a number of challenges including fast recruitment requiring an unusually high frequency of safety reviews, more frequent use of complex designs and virtually no prior experience with the disease.  In this paper, we provide a perspective on the work of DMCs for clinical trials of treatments for COVID-19. More specifically, we discuss organizational aspects of setting up and running DMCs for COVID-19 trials, in particular for trials with more complex designs such as platform trials or adaptive designs. Furthermore, statistical aspects of monitoring clinical trials of treatments for COVID-19 are considered. Some recommendations are made regarding the presentation of the data, stopping rules for safety monitoring and the use of external data. The proposed stopping boundaries are assessed in a simulation study motivated by clinical trials in COVID-19.

## 1. Introduction

The first clusters of Coronavirus disease 2019 (COVID-19) cases were reported in December 2019 and January 2020  [1] [2] [3] [4]. On 11 March 2020, the World Health Organization declared the outbreak of SARS-CoV-2 a pandemic [5]. As of 18 July 2020, over 14 million cases and over 600,000 deaths of COVID-19 were confirmed according to the Center for Systems Science and Engineering at Johns Hopkins University [6] [7].

A search in clinicaltrials.gov for studies targeting the conditions "COVID-19", "COVID", or "SARS-CoV-2" shows that the first studies surrounding COVID-19 were registered in late January 2020 and until July 2020 over 2500 studies were registered. Clinical trials studying interventions for COVID-19 primarily focus on short-term endpoints assessing mortality, morbidity, the requirement for mechanical ventilation or ICU care. For instance, the primary endpoint in the RECOVERY trial (ClinicalTrials.gov Identifier: NCT04381936) is all-cause mortality at 28 days [8], the primary endpoint in the Adaptive COVID-19 Treatment Trial (ACTT; ClinicalTrials.gov Identifier: NCT04280705) was time to recovery within 28 days after enrollment [9], and the primary endpoint in the GS-US-540-5773 trial (ClinicalTrials.gov Identifier: NCT04292899) was the clinical status on day 14, assessed on a 7-point ordinal scale [10].

Well-conducted double-blind randomized controlled trials are considered the gold standard for clinical trials and there have been calls for their rigorous application in COVID-19 [11]. However, conducting a clinical trial for a pandemic disease to established standards in the midst of an evolving pandemic poses a number of challenges [12]. For instance, the location of areas with high numbers of infections changes over time. Therefore, clinical trial sites might need to pause or even stop recruitment which in turn means that new sites have to be opened in different locations. Sites in locations severely affected by the pandemic might be able to screen, randomize and treat a large number of subjects within a short period of time, however, this brings challenges for on-site trial personnel to properly document the cases and enter the data in a timely manner into the study database. Moreover, due to the seriousness of COVID-19, standard of care or best available therapy instead of placebo are included as comparator in many trials, at least as of Summer 2020, but what constitutes standard of care or best available therapy is changing rapidly due to efficacious treatments being identified, e.g. remdesivir [9] [10] or dexamethasone [13]; treatments being granted and then possibly revoked Emergency Use Authorizations (EUA), e.g. hydroxychloroquine sulfate [14] [15]; treatment effects varying based on subjects' health status, e.g. the effect of dexamethasone varying with the respiratory support received at randomization [13].

A Data Monitoring Committee (DMC) is a body, independent of the trial's sponsor, that is tasked with '[...] performing periodic benefit-risk assessments using available efficacy and safety outcomes data gathered during the course of a trial [...]' [16]. In particular, in order to adequately assess the benefits and risks of an intervention, the DMC should have access to all necessary data [17] [18]. Based on their review of the data, the DMC provides recommendations to the trial's sponsor or steering committee to stop the trial early for efficacy or futility, to stop the trial for harm, or to recommend continuation of the trial with or without modifications of the study protocol [19]. Generally, DMCs are comprised of physicians with specialized knowledge of the disease area for which an intervention is studied, and (at least) one statistician. All DMC members should have experience in the conduct of clinical trials and an understanding of a DMC's work [20] [21]. Regulatory guidance regarding DMCs are provided by the FDA, EMA, and the WHO [22] [23] [24].

In this paper, we provide a perspective on the work of DMCs for clinical trials of treatments for COVID-19. In Section 2, we discuss organizational aspects for these DMCs. In Section 3, we focus on statistical aspects of monitoring clinical trials of treatments for COVID-19 including presentation of data for safety reviews, stopping boundaries for safety monitoring and inclusion of external data. We conclude with a discussion of results and limitations in Section 4.

## 2. Organizational aspects

Accrual of subjects in clinical trials of an intervention for COVID-19 is expected to be more rapid than the accrual for clinical trials that study non-pandemic diseases, in particular if the trial is conducted in an area with high numbers of COVID-19 cases. In these instances, the target number of subjects in the clinical trial may be expected to fully recruit within a few months and sometimes even within weeks. A rapid accrual of subjects might require frequent safety monitoring by the DMC with possibly weekly safety reviews. The high DMC meeting frequency is associated with a considerable time commitment by DMC members. Under consideration of the logistics associated with setting up a DMC, such as writing the DMC charter and the DMC SAP, but also ensuring the contractual basis for DMC members' work, it can be efficient to set up the DMC to oversee not only an individual trial but to monitor a program of trials on a disease level. While it is not uncommon that DMCs oversee multiple clinical trials, the oversight is usually for multiple trials studying the same intervention and not multiple trials studying different interventions.

Beyond the operational benefits, a single DMC overseeing multiple studies may then take into account the emerging data from all trials into the decision making. In general, separate studies should not be lumped together and then analyzed as a single study, but different studies should be treated as strata when analyzed jointly [25]. The topic of formalizing the use of external data is discussed further in Section 3.3. The novelty of COVID-19 and the disease's characteristics and progression strongly suggest that a DMC for a clinical trial studying treatments for COVID-19 should be multi-disciplinary, beyond the two clinicians and one statistician that traditionally form a DMC. For instance, in addition to a pulmonologist, the DMC may include a physician with expertise in intensive care medicine if the study focuses on hospitalized COVID-19 subjects. With a wide range of drugs being administered to patients with COVID-19, a pharmacologist's expertise brings additional value to the DMC. Last but not least, a clinical epidemiologist and an infectious diseases expert provide relevant knowledge to the DMC for a clinical trial of COVID-19 treatments.

The quality of data available for the DMC review may be affected by the accrual speeds and the data may not be of equivalent quality that is generally provided during data reviews. For example, due to the fast accrual of subjects, sites may not have the personnel required for a timely entry of data into a database, or entered data for a subject may be incomplete. For example, a conceivable scenario is that a serious adverse event is reported, but that for the same subject data on concomitant medications or the medical conditions are not available to the DMC at the time. Incomplete data may complicate the committee's ability to draw adequate conclusions from the provided data. Therefore, the closed report provided to the DMC should include measures assessing the availability, quality and completeness of the data, as it should be the usual custom [26] [27]. In addition, the DMC should be made aware of any incomplete records, errors and general inconsistencies in the data. Informative reporting on the quality of data becomes considerably more important when the quality of data may affect the interpretation of the data.

For COVID-19, there are a number of platform or multi-arm trials including the World Health Organization's Solidarity trial [28], the RECOVERY trial (ClinicalTrials.gov Identifier: NCT04381936) [8] and the ACTT trial (ClinicalTrials.gov Identifier: NCT04280705) [9]. Furthermore, the use of efficient adaptive designs in COVID-19 has been advocated and implemented in some ongoing trials [29]. Whereas adaptations based on blinded (i.e. non-comparative) data usually do not require a DMC, adaptations based on unblinded (comparative) data do require an independent party, typically the DMC. Therefore, the responsibilities of the DMC go beyond the standard safety monitoring, since the DMC will review unblinded (comparative) data to make recommendations regarding preplanned adaptions, such as treatment selection in multi-arm trials, subgroup selection or sample size re-estimation. Alternatively, a separate committee advising on the adaptations, sometimes referred to as Adaptation Committee, may be setup in addition to the DMC for safety monitoring. To our knowledge, however, this model is not common in clinical trial practice.

If the DMC is charged with the responsibilities regarding the adaptations, the DMC needs to have the necessary expertise and experience with adaptive designs. Furthermore, the interactions between varies parties including the DMC and the sponsor, in particular the clinical trial team, are more complex in this type of trials. The DMC makes recommendations and the sponsor decides whether or not to follow these. Typically these are followed, but with some exceptions (see for instance Filippatos et al. [30]). In adaptive designs, the chances that the recommendation may not be followed may be higher than in standard designs where the DMC is only concerned with safety monitoring, since some adaptations may have consequences for the labeling und ultimately marketing and clinical use when, for example, adaptations concern dose or subgroup selections. This means in turn that there must be an opportunity for discussions between the DMC and the sponsor to resolve such matters without unblinding the clinical trial team. This is achieved by installing sponsor representatives (also sometimes referred to as sponsor committee). Typically, the sponsor representatives are sponsor personnel with the required expertise and seniority to make the necessary decisions, but who are at the same time independent of the clinical trial team [31]. With such sponsor representatives in place, the DMC submit their recommendations to them rather than the clinical trial team.

## 3. Statistical aspects of monitoring clinical trials in COVID-19

In this section, we consider some statistical aspects of monitoring clinical trials in COVID-19. We start by discussing the scope of the data reviewed by the DMC and their presentation. Then stopping rules for safety monitoring are considered and evaluated in a simulation study. We briefly describe how external data may be incorporated in safety reviews and formal interim analyses before reflecting upon some particular issues that may arise with more complex designs, such as platform trials or adaptive designs.

### 3.1. Scope and presentation of data

The report provided to the DMC needs to enable the DMC to get a comprehensive understanding of the intervention's safety, efficacy and benefit-risk profile [32]. The details of DMC reports are intervention and trial specific, but common content include baseline characteristics, participant disposition, treatment exposure, protocol adherence, safety data, lab values, and efficacy data [27]. Providing the DMC with reports on the quality of data is important, too, particularly when the accrual rate may affect the data quality, as discussed in Section 2. Data quality measures include, but are not limited to, the number of subjects who were randomized and treated, who completed the study, who withdrew consent, who stopped treatment due to adverse events, and the delay between data being collected and reported. For clinical trials assessing interventions for COVID-19, the comparator in many trials is the standard of care at least as of the time of submission (August 2020). With the standard of care changing rapidly over time, and possibly differing between regions, countries, and even between sites, information on the standard of care and concomitant medications should be presented stratified by the relevant location.

DMC reports should be have a clear structure and ideally be a single document that includes a table of content. The graphical and interactive visualization of data may ease the exploration of the data and enhance the readers' understanding of the data [33] [34] [35]. DMC reports are no exception to this. Examples for DMC reports that fulfill the previously described principles are for instance provided by the University of Wisconsin-Madison [36] [37]. In addition, Evans et al. [32]

recently recommended to show forest plots of risk differences/ratios for key safety and efficacy endpoints, plots of rates of ranked desirable outcomes (DOOR plots), or so-called lasagna plots that summarize the benefit-risk over time by treatment.

In addition to a report, the DMC may review data during their meeting using interactive displays. Interactive displays can for example be created with Shiny, an R package for building interactive web apps from R. Open-source code implementing an interactive display of adverse event data is available in JavaScript [38] [39] and R [40]. Moreover, an R package for creating interactive graphic for clinical trial safety data has been published [41]. An interactive review of data through apps may simplify the data review process and results in a more comprehensive of the data. Let us consider an interactive display of adverse event data as an example. Such an interactive display may start with a comparison of the number of AEs by primary system organ class between groups that provides a high level overview. Through filters for seriousness, severity, or relationship to the treatment, relevant information may be displayed immediately. By linking the number of events to a list of the subjects' IDs that experienced these events and then linking the IDs to additional event information, lab parameters, etc, the DMC may be provided all information relevant to their safety assessment instantaneously. A screenshot of such an interactive display of adverse event data is shown in Figure 1. The versatile tools facilitates the exploration of the data on a group and individual level.

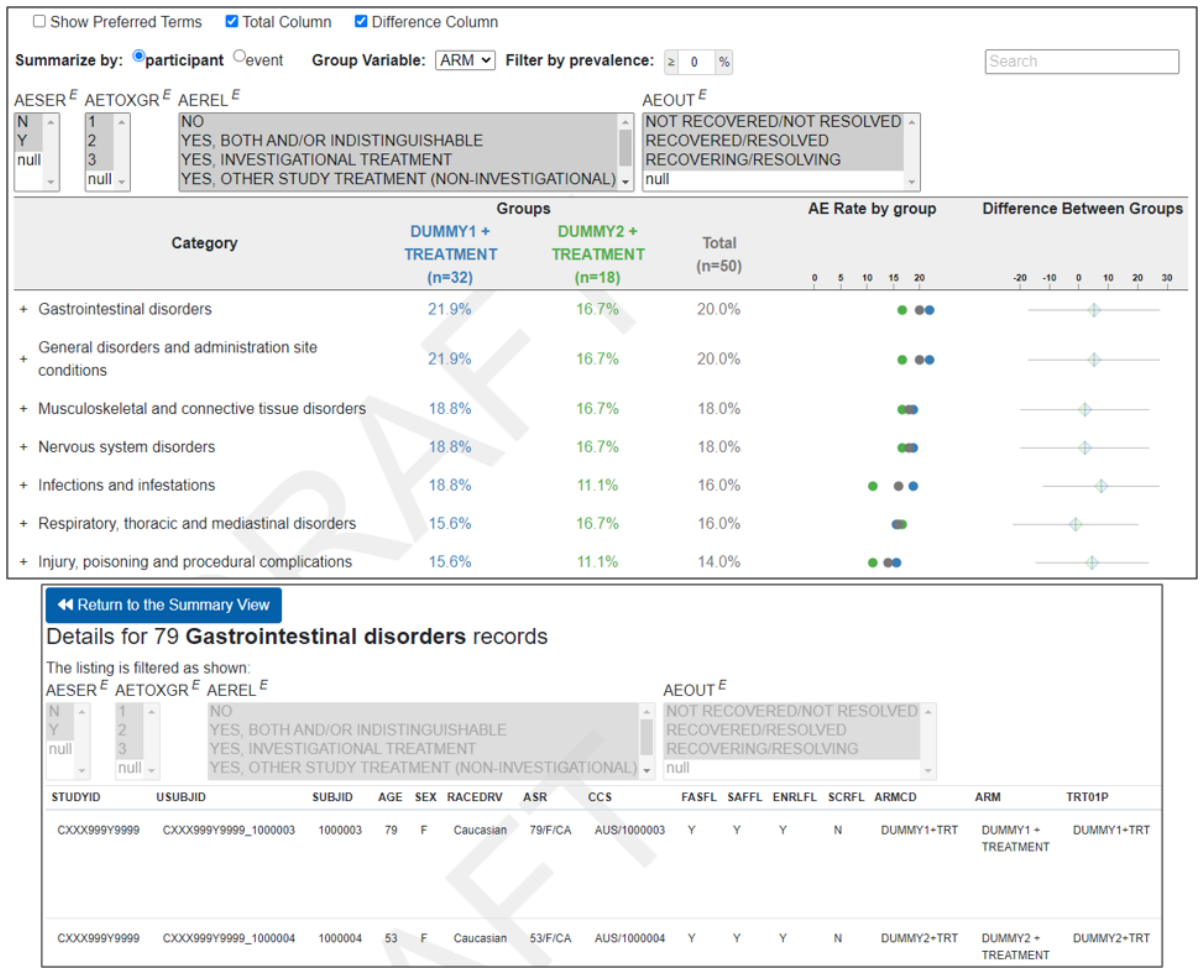

*Figure 1 Screenshots of an interactive display of adverse event data. Top: interactive display of the comparison of adverse event rates between group by system organ class is shown. Bottom: Details of the subjects for whom a gastrointestinal disorder was reported. The details are obtained by clicking on 'Gastrointestinal disorder' in the interactive display shown on top.*

An example for an interactive display of laboratory data is shown in Figure 2. The display allows the selection of the laboratory value of interest and then plots for each subject the measurement against the time point. The y-axis limits and the method for highlighting the 'normal' range can be selected by the user. By clicking on one of the trajectories for a subject, the trajectory is highlighted in bold and additional graphical displays of laboratory measures for that subject are shown.

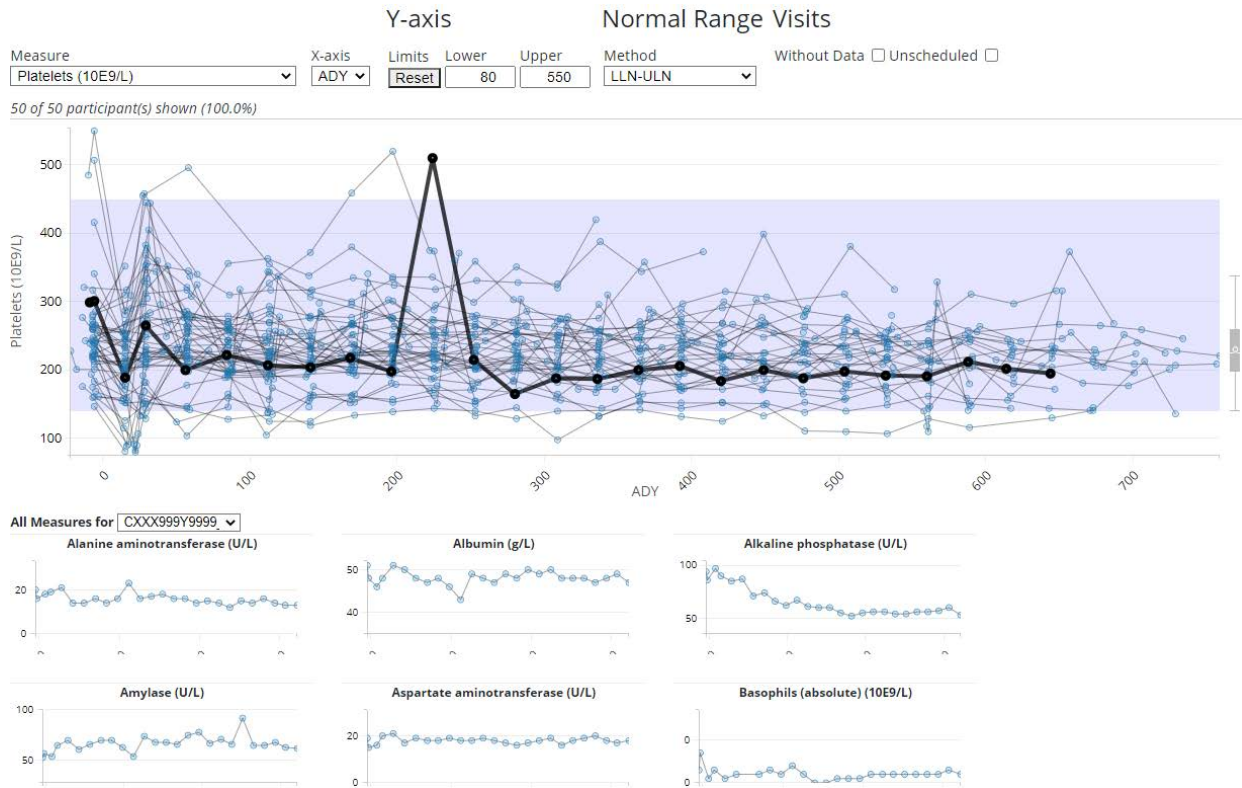

*Figure 2 Interactive display of laboratory values (here: Platelet counts).*

## 3.2. Monitoring of time-to-event data for safety assessments

All-cause mortality, mechanical ventilation, and transition to ICU care are important events that should be monitored by the DMC as part of their safety assessments of an experimental treatment for COVID-19. Death prevents the occurrence of further events, e.g. mechanical ventilation or ICU case. In other words, the events are of competing nature. This complicates the monitoring, because fewer subjects may require mechanical ventilation in one group compared to the other, but this difference may be caused by an excess of early deaths. Appropriate statistical methodology taking into account the competing nature of the events needs to be employed by the DMC when monitoring the events [42] [43] [44] [45] . An alternative to monitoring the events separately is to monitor the composite event, that is event-free survival. A sensible approach to selecting the monitoring guidance is studying the operating characteristics of different rules and selecting the one which has the most desirable properties under a range of realistic assumptions. Operating characteristics of interest are the probability of detecting harm and the probability of erroneously stopping the trial for harm. Both probabilities have to be balanced when deciding on a monitoring guidance. Unlike efficacy analyses where the overall probability of erroneously stopping for controlled at the one-sided level of 2.5%, there is no agreed threshold for safety analyses. In the following, we describe the process for selecting a monitoring guidance for a single time-to-event variable, e.g. death or mechanical-ventilation-free survival. It is established practice to specify such guidance as non-binding in the DMC charter such that when the monitoring boundary is crossed, the DMC has no obligation to recommend stopping the study, but may take the totality of data into account in their recommendation [46].

Let $i = 1, \ldots, n$ index the subjects in the trial and let $X_i$ be a subject's treatment variable which is $X_i = 1$ for subjects in the experimental treatment group and $X_i = 0$ for subjects in the control group. To monitor the event, we employ the Cox proportional hazard model with the treatment variable as a fixed factor [47]. The hazard function is given by

$$\lambda(t|X_i) = \lambda_0(t)\exp(X_i\beta).$$

The hazard ratio $HR = \exp(\beta)$ may be estimated by maximizing the partial likelihood. A hazard ratio greater than 1, $HR > 1$, corresponds to the experimental treatment causing harm and a hazard ratio smaller than 1, $HR < 1$, corresponds to a protective effect of the experimental treatment. Then, the question of safety versus harm of the experimental treatment (with respect to the monitored time-to-event variable) may be formulated as a statistical hypothesis testing problem

$$H_0: HR = 1 \ \ vs. \ \ H_1: HR > 1.$$

At each monitoring time point, the null hypothesis may be tested with a statistical hypothesis test. A crucial point is the selection of the significance level for the hypothesis test. When monitoring efficacy and futility, group sequential boundaries are generally considered to control the type I error rate of the trial at a level of $\alpha$ [48]. Group sequential boundaries may also be applied to monitoring harm, however, while these boundaries control the probability of erroneously stopping for harm, they generally have low power to detect existing harm, especially early in the trial [49]. Alternatively, to increase the probability that the monitoring procedure detects existing harm, the test for $H_0$ at each look may be performed with a nominal significance level of $\alpha$, e.g. $\alpha = 0.025, 0.05, 0.1$.

Next, we present a simulation study motivated by settings typical for COVID-19 trials. In the simulations, the operating characteristics of a monitoring-for-harm procedure for a time-to-event variable based on the Cox regression are assessed. We consider a two-arm clinical trial where subjects are followed up for four weeks, i.e. 28 days, since this is typical for COVID-19 treatment trials. The target sample sizes are $n = 500, 1000$ and the treatment allocation is 1:1. We focus on the settings in which the recruitment is uniform over a period of eight weeks. The events are simulated using an exponential distribution. The event rate is chosen such that a subject on control experiences an event within the four weeks follow-up with a probability of $P(\text{Event within 4 weeks|CTL}) = 0.15$. On the experimental treatment, the probability of experiencing an event is varied between 0.15 and 0.25, i.e. $P(\text{Event within 4 weeks|TRT}) = 0.15, 0.175, 0.2, 0.25$. To monitor for harm, a test of $H_0: HR = 1$ with a one-sided significance level for $\alpha = 0.025, 0.05$ based on the Cox regression is performed at each data look. The monitoring is conducted on a weekly basis starting one week after the randomization of the first subject. Based on the probabilities that an event occurs within four weeks after randomization in the treatment group and the control group, the hazard ratios in the Cox model may be calculated.

*Table 1 Specifications for the simulation study of the monitoring procedure's operating characteristics.*

| Parameter | Value |
|---|---|
| Uniform recruitment period | 8 weeks |
| Sample size | n=500, 1000 |
| Treatment allocation | 1:1 |
| $P(\text{Event within 4 weeks|CTL})$ | 0.15 |
| $P(\text{Event within 4 weeks|TRT})$ | 0.15, 0.175, 0.2, 0.25 |
| Hazard ratio (as a result of assumptions above) | 1, 1.18, 1.37, 1.77 |
| Significance level $\alpha$ | 0.025, 0.05 |

| Monitoring frequency | Weekly |
|---|---|

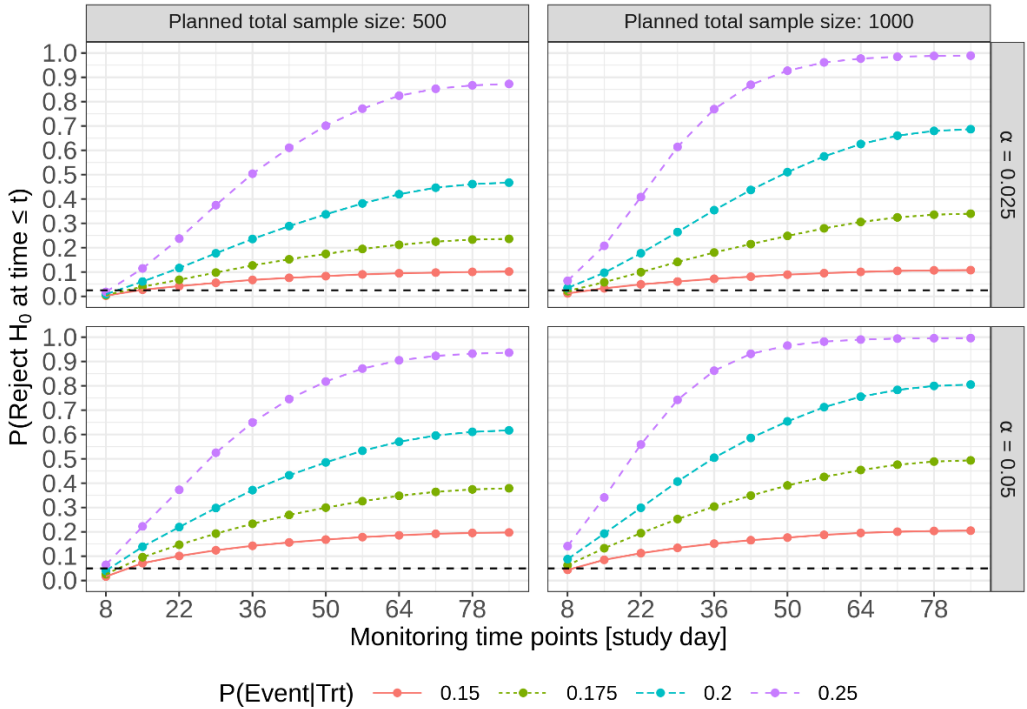

*Figure 3 Cumulative probability for rejecting $H_0$ versus the monitoring time points. At each time point the test is performed with a one-sided significance level $\alpha$.*

Figure 3 shows the probability for rejecting the null hypothesis $H_0$ in favor of the one-sided alternative hypothesis $H_1$ prior to or at monitoring time point $t$. The results are presented for two planned total sample sizes, four different probabilities of experiencing an event within the four weeks follow-up under treatment, that is $P(\text{Event within 4 weeks}|\text{TRT})$, and two one-sided significance levels $\alpha$. The red line shows that due to the repeated testing of the null hypothesis $H_0$ at the one-sided significance level $\alpha$, the cumulative probability to wrongly reject the null hypothesis during at least one monitoring time point increases to about 0.1 for $\alpha = 0.025$ and to 0.2 for $\alpha = 0.05$. Figure 3 also shows that probability to detect differences in the event rate between the treatment group and the control group increases with the sample size and that larger differences are naturally easier to detect. Moreover, the probability to detect differences between the groups increases with increasing significance level $\alpha$, but the probability to wrongfully detect differences also increases.

It is worth highlighting that monitoring harm by testing the null hypothesis at each time point with a fixed significance level is a group sequential design with a Pocock boundary for which not the global type I error rate but the significance level at each time point is chosen [50]. Therefore, a large sample approximation of the simulation results presented in Figures 1 and S1 can be obtained through standard group sequential software such as the R package *gsDesign* [51]. In detail, the cumulative probability for rejecting $H_0$ at or prior to the time point $t$, that is

$$P(Reject\ H_0\ at\ time \leq t) = P(T_k \geq q_{1-\alpha}\ for\ any\ k \in \{t_1, t_2, \ldots\}),$$

with $t_1, t_2, \ldots$ the monitoring time points, $q_{1-\alpha}$ the $(1-\alpha)$-quantile of a standard normal distribution, and $T_k$ the Wald statistic for testing $H_0$. The joint distribution of test statistics $T_k$ can be approximated by a multivariate normal distribution where each component has mean $\beta\sqrt{m_k r_T r_C}$ and variance one [52] [53]. Here, $m_k$ is the expected number of events at time point $k$, and $r_T$ and $r_C$ is the proportion of subjects in the treatment and control group, respectively. The correlation of test statistics from time points $t_1 < t_2$ is approximated by $\sqrt{m_{t_1}/m_{t_2}}$. This normal approximation can then be used to calculate the cumulative probability for rejecting $H_0$ at or prior to the time point $t$ [48]. Figure 4 shows that the approximation is satisfactory, in particular for the setting with total sample size of $n = 1000$.

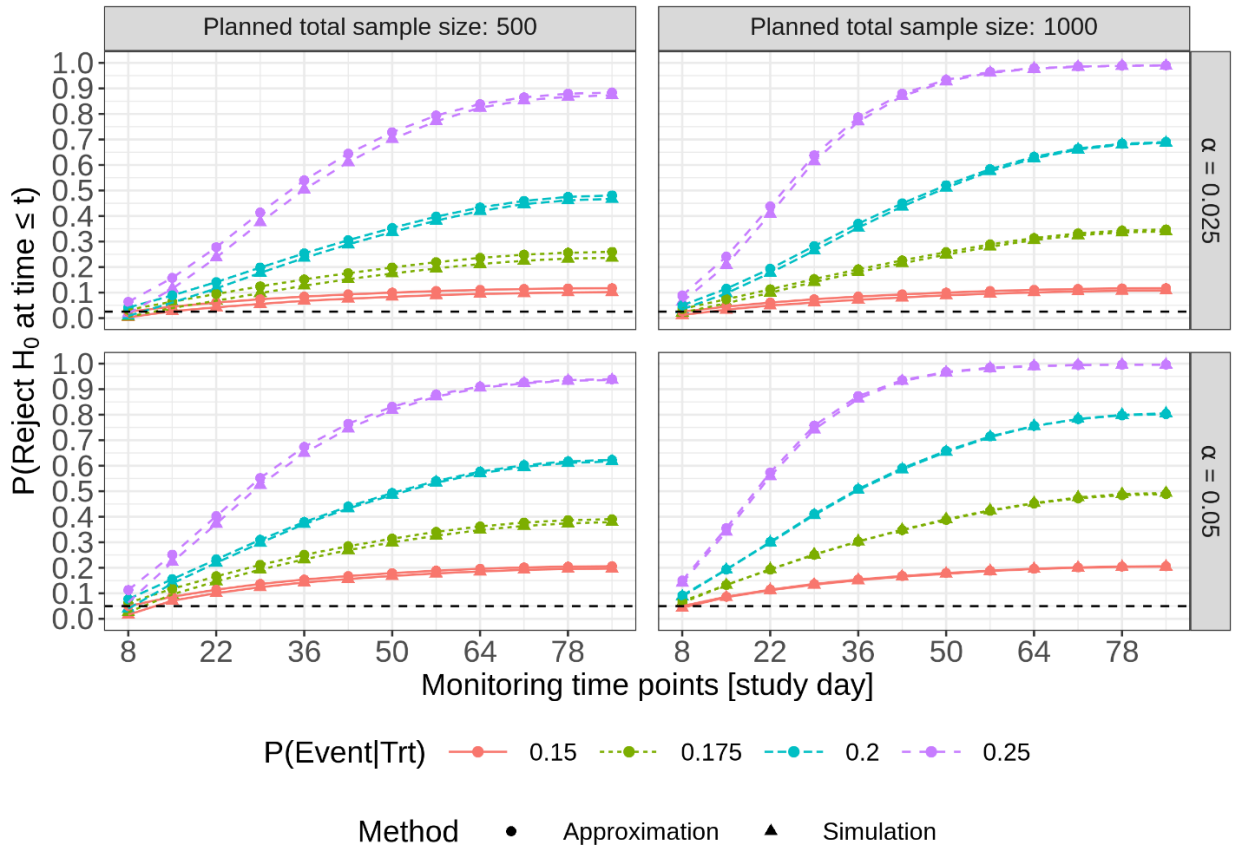

*Figure 4 Cumulative rate for rejecting $H_0$ versus the monitoring time points based on simulations (as in Figure 3) and on the asymptotic normal approximation. The ratio of the expected number of events from different monitoring time points is considered as the correlation of test statistics the corresponding time points. At each time point the test is performed with a one-sided significance level α.*

## 3.3. Incorporating external data

A DMC does not consider data from the trial monitored in isolation, rather data in the context of other available or emerging data. We refer to any data outside the monitored trial as external data. These may be from randomized controlled trials or other types of studies including clinical registries. In particular, in situations of rapidly changing external landscapes such as with COVID-19, a DMC must be aware of any new safety or efficacy signals that may arise for the same drug or drugs with similar mechanisms of action. In COVID-19, there were a number of trials ongoing assessing the efficacy and safety of hydroxychloroquine. The perception of hydroxychloroquine changed quite dramatically of the course of only a few weeks. At first it was considered a promising treatment option, then suspected to be unsafe and finally dismissed for lack of efficacy [15].

If there is agreement that external data should be included, the question remains how this could be achieved. In principle, the evidence could be included informally, e.g. by considering data side by side but not combining them statistically, or formally, e.g. by using meta-analytic approaches [54]. One critical point in combining data is the similarity of the monitored trial and the studies providing the external evidence in terms of study design, patient population, standard of care etc. When integrating the data formally, e.g. through a random-effects meta-analysis, this will be capture in the between-trial heterogeneity. In the following we make some recommendations on the formal integration of external evidence with regard to adverse events [45].

Unfortunately, it is still common to pool adverse event data naively across studies by "simply combin[ing] the numerator events and the denominators for the selected studies" [55], although this might lead to bias due to Simpson's paradox [56] [57] [58]. Therefore, the use of meta-analysis techniques is encouraged. These may account for heterogeneity in the control group outcomes across studies and, if random-effects meta-analysis is used, also in treatment differences. A number of problems are faced with safety analyses (see, e.g. [59]). These include varying follow-up times between studies, rare events and small numbers of studies included in the meta-analysis. The latter makes estimates of the between-study heterogeneity in the treatment differences uncertain with negative consequences for the inference regarding the overall treatment effect [60]. Bayesian approaches using weakly informative priors for the between-study heterogeneity have been suggested to deal with this problem in the normal-normal hierarchical model, the standard model for random-effects meta-analysis [61]. The application of such techniques is straightforward using the R package *bayesmeta* available on CRAN [62]. Furthermore, they may be extended for applications with rare events using in addition also a weakly informative prior on the treatment effect [63].

When combining the evidence from the monitored trial with external evidence the primary interest might not be in the overall effect but rather in the effect of the monitored trial in the light of the external evidence, the so-called shrinkage estimate [64]. In a Bayesian framework, this may be understood as using the posterior of a meta-analysis as the prior for the analysis of the new study, the so-called meta-analytic predictive (MAP) prior approach [65]. If the analyses are carried out in the normal-normal hierarchical model the shrinkage estimates are included in the standard output of the *bayesmeta* package.

The methods discussed so far are applicable if the same quantity is observed in all studies. For instance, this may be a treatment contrast or an event probability. Following suitable transformations such as the logarithmic transformation for e.g. relative risks or hazard ratios these quantities may be combined in a meta-analysis and the overall effect or shrinkage estimates derived. In some cases where different quantities are observed for all studies, the estimates of the treatment contrast of interest might not be available from other studies, but only data on the control group. Hence, a variation of the MAP approach may be used to summarize external evidence on the control group and combine it with the control of the monitored trial [66]. In practice, such approaches can be implemented using the R package *RBest* [67].

## 3.4. Complex designs

As discussed in Section 2, some of the interventional trials in COVID-19 use more complex designs including platform trials or adaptive designs. A review of the methodology as well as recommendations and examples may be found for instance in the recent paper by Stallard et al [29]. They also provide a comprehensive list of references on methods and applications of adaptive designs. Here we briefly comment on some aspects relevant to DMCs.

The longer the accrual period in relation to the follow-up period (or vice versa the shorter the follow-up period in relation to the accrual period) the more favorable is the situation for adaptive designs [68]. Although recruitment in some COVID-19 trials is quite fast (e.g. ACTT recruited more than a 1,000 patients in less than three months), adaptive designs may still be applied since the endpoints are also observed quite quickly with follow-up periods of up to 4 weeks commonly used in COVID-19 trials (see examples provided in Section 1). Furthermore, adaptations such as treatment or subgroup selection might also be based on early outcomes, e.g. shorter term readouts of the final outcome [69]. To plan such trials the R package *asd* is available from CRAN [69] [70].

Regarding multi-arm trials there has been some debate with regard to the control of error probabilities. Although some authors suggest to control the familywise type I error rate (FWER) at the trial level (see e.g. [71] [72]), others argue that this should not be the default for clinical trials evaluated distinct treatments [73]. The main argument against control of the FWER is it not being controlled if the treatments were assessed in separate trials. This discussion has been reflected upon in a regulatory setting and extended to master protocols [74].

Whether or not to recommend stopping of a trial either for futility or early success is often a complex and difficult decision to make. As the recent example of ACTT shows, decisions to unblind a trial early might spark some discussions [75]. In brief, ACTT compared remdesivir with placebo enrolling 1062 patients. The primary endpoint was the time to recovery. Following some promising results observed in an interim analysis by the DMC (in the trial referred to as the data safety monitoring board) the data were made public and patients on placebo could receive treatment with remdesivir. However, the trial did not demonstrate any statistically significant benefit in mortality [9]. Therefore, the ultimate proof of efficacy is still missing and one may only speculate on the outcome of the trial with longer follow-up or larger sample size.

## 4. Discussion

In the unfolding SARS-CoV-2 pandemic, DMCs of interventional trials face a challenging task as there is on the one hand some pressure for early termination of trials due to the unmet need for treatments and on the other hand the need to meet established standards for the evaluation of treatments in terms of efficacy, safety, and ultimately benefit-risk. Furthermore, the understanding of COVID-19 is evolving and at least at the outset of many trials, not well understood. In addition, the accrual period is shorter than in comparable trials in intensive care settings requiring frequent safety reviews. Here, we discussed a number of logistical and statistical aspects of DMCs for COVID-19 interventional trials. In particular, we recommended a safety monitoring rule that might also proof to be useful in other diseases. The rule uses the hazard ratio of a time to event outcome such as all-cause mortality or event-free survival and indicates stopping the trial for safety concerns if the hazard ratio is nominally significant at a pre-specified level. The rule was evaluated in a simulation study motivated by ongoing trials in COVID-19. Furthermore, we demonstrated the interpretation of that rule as group sequential Pocock stopping boundaries at an elevated significance level.

For early clinical trials, the use of internal DMCs is common. These are independent of the clinical trial team but not external to the sponsor and therefore not independent of the sponsor. In particular for registration studies, however, the standard is that DMCs are generally comprised of members that are not only independent of the study team, but also independent of the sponsor to ensure the integrity and validity of the trial [30]. Given the time pressure and the difficulties of setting up DMCs during an ongoing pandemic, one might consider DMCs that are not purely external but include some of the sponsor's internal expertise and experience.

Here we discussed the role of DMCs in COVID-19 trials. However, DMCs have of course an important role to play in trials in non-COVID-19 diseases that are impacted by the SARS-CoV-2 pandemic. However, they should not be involved in decisions on any design changes if they have been exposed to unblinded (comparative) data [76] [77].

## Acknowledgements

The authors would like to thank Douglas M. Robinson for comments on an earlier version of the manuscript and the Scientific Computing and Consulting group for providing the Shiny apps which are shown in screenshots in this manuscript.